\documentclass[nofootinbib,11pt,showpacs]{revtex4-1}

\usepackage{epsfig}
\input epsf.tex
\usepackage{amssymb}
\usepackage{amsmath}
\usepackage{amsthm}
\usepackage{amsopn}

\def\comment#1{}

\def\beq{\begin{equation}}
\def\eeq{\end{equation}}
\def\bea{\begin{eqnarray}}
\def\eea{\end{eqnarray}}

\usepackage{ulem}
\usepackage{cancel}
\usepackage{color}

\begin{document}
{\textsf{\today}
\title{
Neutrinos as a probe of curvature
 }
\author{  Jafar Khodagholizadeh }
\affiliation{Farhangian University, P.O. Box 11876-13311, Tehran, Iran.\\
	Gholizadeh@ipm.ir
}

\begin{abstract}
	Neutrinos, as the anisotropic stress tensor, have a damping effect on the tensor mode perturbation from inflation to the $ \Lambda$ dominated era. First, we study the squared amplitude reduction for the wavelength entering the horizon during radiation and matter-dominated phases in the negatively curved de Sitter spacetime. Then, by comparing with other spatial spacetimes, $ K=0 $ and $K=1$, the highest difference between closed and open cases is seen in the matter-dominated era. Thus, neutrinos can be added as another candidate for determining the nature of space-time.
\end{abstract}

\maketitle

\section{Introduction}
The exact shape of the universe is still a matter of debate in cosmology. Although experimental data from various independent sources such as WMAP and Planck confirm that the universe is flat and infinite \cite{1}, there is deeper evidence that is consistent with the other possible shapes.\\
Recently, Planck 2018 CMB lensing \cite{lensin} and baryon acoustic oscillations \cite{14, 15, 16} suggest a flat universe is quantitatively inconsistent at $ 2.5 $ to $ 3\sigma $  with CMB data. Besides, the Planck legacy, 2018 CMB power spectra provides statistically significant indications for a closed universe \cite{Pl18}. The curvature tension and ages of the oldest stars are considered pieces of evidence for the closed spacetime\cite{Will,Schla,Bond}.\\
On the other hand, there is the opposite scenario which is consistent with the negatively curved FRW models: Gamma-ray halos around active galactic nuclei \cite{Ando:2010rb}, the intergalactic magnetic field with the observation of TeV blazar 1ES 0229+200 \cite{Tavecchio:2010mk}, gamma-ray data \cite{Essey:2010nd}, and strong magnetic filed from Fermi observations of TeV blazars\cite{Neronov:1900zz}. Also the quintessence models with various potentials   \cite{Aurich:2001fe}, and a nonlinear massive gravity with Lorentz-invariant mass terms  have been studied  in open FRW-Universe \cite{Gumrukcuoglu:2011ew, Dariescu:2016udv}.\\
In general, tensor perturbations arise from inflationary in the early universe and propagate freely in the expanding universe, which is a perfect fluid \cite{Lifshitz}. It was thought that the interaction of these modes with matter and radiation could be negligible \cite{Mukhanov}, but Steven Weinberg showed far outside the horizon at the temperature $ \approx 10^{10~0} K$ tensor modes can be affected by anisotropic inertia which contain free streaming neutrinos and antineutrinos \cite{damping}. This issue was explored in the spatially closed spacetime and extended to the $ \Lambda$-dominated era \cite{Khodagholizadeh:2014ixa, Khodagholizadeh:2017ttk}.
Here, in addition to the above reports, we will use the neutrinos effect on gravitational waves damping as a tool to show the difference between spatially closed and open universe. \\
To calculate this effect in spatially open de Sitter, this paper is organized as follows.  In Sec. II, we review the relativistic Boltzmann equation for the number density of neutrinos in the presence of tensor perturbation in the generally curved spacetime. In Sec. III, we present the integrodifferential equation of gravitational waves with neutrinos in the spatially open de Sitter spacetime. In the next sections, we derive the general solutions of GWs with damping effects in radiation, matter, and expanding universe. In the last section, we will discuss the issue by comparing our results with the flat and closed cosmology.
\section{DAMPING source IN THE WAVE EQUATION}
The tensor fluctuation in curved spacetime satisfies as \cite{gholizadeh}
\begin{eqnarray}\label{4}
\nabla^{2}D_{ij}-a^{2}\ddot{D}_{ij}-3a\dot{a}D_{ij}-2KD_{ij}=16\pi G\Pi_{ij}
\end{eqnarray}
where the dot stands for derivative concerning the ordinary time, the tensor mode gravitational perturbation takes the form:
$ \delta g_{ij}=h_{ij}(\vec{x},t)=a^{2}(t) D_{ij}(\vec{x},t)$,  $ K $ is curvature constant, and $ \Pi_{ij} $ is anisotropic inertia tensor which is the sum of the contributions from photons and neutrinos; but photons have small contribution due to the short mean free time. Neutrinos (including antineutrinos) travel without collisions when the temperature drops about  $ T=10^{10} K $, the distribution function of which in the phase space has the form:
\begin{equation}
n_{\nu}(\vec{x},\vec{p},t)\equiv \sum_{r}\prod_{i=1}^{3}\delta^{(3)}(x^{i}-x_{r}^{i}(t))\prod_{i=1}^{3}\delta^{(3)}(p_{i}-p_{ri}(t))
\end{equation}
where $  r $ labeling trajectories of individual neutrinos (or antineutrions) which $ n_{\nu}(\vec{x},t) $ in the start of free streaming has the form of the ideal gas as: $
\bar{n}_{\nu}(\vec{x},\vec{p},t)=\dfrac{N}{(2\pi)^{3}}/[exp(\frac{\sqrt{g^{ij}p_{i}p_{j}}}{k_{B}a(t)\bar{T}(t)})+1]
$ where $ N $ is the number of types of neutrinos and separately antineutrinos,
and $ k_{B} $ is the Boltzmann constant. In small perturbation to the metric, the neutrino distribution function is slightly different from its equilibrium form as:
$
n_{\nu}(\vec{x},\vec{p},t)=n_{\nu}(a(t)\sqrt{g^{ij}p_{i}p_{j}})+\delta n_{\nu}(\vec{x},\vec{p},t)
$. Therefore, in the absence of collision terms, 
the relativistic Boltzmann equation for the perturbation $ \delta n_{\nu}(\vec{x},\vec{p},t) $ in curved spacetime will be:
\begin{eqnarray}\label{2}
\dfrac{\partial\delta n_{\nu} (\vec{x},\vec{p},t)}{\partial t}+\dfrac{p^{i}}{a(t)p}\dfrac{\partial\delta n_{\nu} (\vec{x},\vec{p},t)}{\partial x^{i}}+K \dfrac{\hat{p}_{i}}{a(t)}p x^{l}\hat{p}_{l}\dfrac{\partial\delta n_{\nu}(\vec{x},\vec{p},t)}{\partial p_{i}}=\dfrac{p}{2}\hat{p}_{i}\hat{p}_{j}\bar{n}_{\nu}^{\prime}(p)\dfrac{\partial}{\partial t}D^{ij}(\vec{x},t) ~~~~~~~~~~~~~~~~\nonumber\\
-K \dfrac{\bar{n}_{\nu}^{\prime}(p)}{a}p \hat{p}_{k}x^{m}D^{km}(\vec{x},t)-K \dfrac{\bar{n}_{\nu}^{\prime}(p)}{a}p (x^{l}\hat{p}_{l})\hat{p}_{i} \hat{p}_{k}D^{ki}(\vec{x},t)+K^{2} \dfrac{\bar{n}_{\nu}^{\prime}(p)}{a}p x^{i}(x^{l}\hat{p}_{l})^{2}\hat{p}_{k} D^{ki}(\vec{x},t)\nonumber\\ -K^{2} \dfrac{\bar{n}_{\nu}^{\prime}(p)}{2a(t)}p \hat{p}_{i}x^{j}x^{k}(x^{l}\hat{p}_{l})^{2}\dfrac{\partial}{\partial x^{i}}D^{jk}(\vec{x},t)+K^{3} \dfrac{\bar{n}_{\nu}^{\prime}(p)}{2a} x^{i}x^{j}x^{k}(x^{l}\hat{p}_{l})^{3}\dfrac{\partial}{\partial x^{i}}D^{jk}(\vec{x},t)\nonumber\\
\end{eqnarray}
With a dimensionless intensity perturbation $ J $, defined by:
\begin{eqnarray}
a^{4}(t) \bar{\rho}_{\nu}(t) J(\vec{x},\vec{p},t)\equiv N_{\nu}\int_{0}^{\infty}\delta n_{\nu}(\vec{x},\vec{p},t) 4\pi p^{3}dp
\end{eqnarray}
where $ N_{\nu} $ is the number of species of neutrinos and antineutrinos, and $ \bar{\rho}_{\nu} \equiv N_{\nu} a^{-4}\int 4\pi \rho^{3}\bar{n}_{\nu}(p)dp$. Thus the Boltzman equation becomes 
\begin{eqnarray}\label{5}
\dfrac{\partial}{\partial t} J(\vec{x},\vec{p},t)+\dfrac{p^{i}}{a(t)p}\dfrac{\partial}{\partial x^{i}} J(\vec{x},\vec{p},t)-4 K \dfrac{x^{l}\hat{p}_{l}}{a(t)} J(\vec{x},\vec{p},t)=-2\hat{p}_{i}\hat{p}_{j}\dfrac{\partial}{\partial t}D^{ij}(\vec{x},t) +\dfrac{4K}{a(t)}\hat{p}_{k}x^{m} D^{km}(\vec{x},t)\nonumber\\
+\dfrac{4K}{a(t)}\hat{p}_{m}\hat{p}_{k} x^{l}\hat{p}_{l}D^{km}(\vec{x},t)- \dfrac{4K^{2}}{a(t)}(x^{l}\hat{p}_{l})^{2} x^{i}\hat{p}_{k}D^{ki}(\vec{x},t)\nonumber\\
+\dfrac{2K^{2}}{a(t)}\hat{p}_{i}x^{j}x^{k}(x^{l}\hat{p}_{l})^{2}\dfrac{\partial}{\partial x^{i}}D^{jk}(\vec{x},t)-\dfrac{2K^{3}}{a(t)}x^{i}x^{j}x^{k}(x^{l}\hat{p}_{l})^{3}\dfrac{\partial}{\partial x^{i}}D^{jk}(\vec{x},t) \nonumber\\
	\end{eqnarray}
The above equation can be have a solution in the following form:
\begin{eqnarray}
J(\vec{x},\hat{p},t)&=&\sum_{\lambda=\pm}\sum_{q}e^{i\vec{q}.\vec{x}}\beta (\vec{q},\lambda)e_{ij} (\hat{q},\lambda)\hat{p}_{i}\hat{p}_{j}\Delta_{\nu}^{T}(q,\hat{p}.\hat{q},t) 
\end{eqnarray} 
and also we define $D_{ij}(\vec{x},t)$ as
\begin{eqnarray}D_{ij}(\vec{x},t)=\sum_{\lambda=\pm}\sum_{q} e^{i\vec{q}.\vec{x}} \beta (\vec{q},\lambda)e_{ij} (\hat{q},\lambda)D_{q}(t)
\end{eqnarray}
where $\beta (\vec{q},\lambda)$ is the stochastic parameter for the single mode with discrete wave number $q$ and the helicity $ \lambda$ and  $e_{ij} (\hat{q},\lambda)$ is the corresponding polarization tensor. Also $\Delta_{\nu}^{T}(q,\hat{p}.\hat{q},t)$ is the helicity-independent function.\footnote{The Botzmann equation for the photon density matrix perturbation, $ \delta n^{ij}$ is generated in terms of temperature $\Delta_{T}^{(T)}$ and polarization $\Delta_{P}^{(T)}$ which superscript $T$ stands for " tensor". Here $\Delta_{\nu}^{T}(q,\hat{p}.\hat{q},t)$ has the role of these terms for neutrinos density matrix perturbation.}
With $ \hat{p}_{i}\hat{q}_{i}=\mu $ the  Eq.(\ref{5}) becomes an equation for $ \Delta_{\nu}^{T} $ as:
\begin{eqnarray}\label{12}
\dfrac{\partial}{\partial t}\Delta_{\nu}^{T}(q,\mu,t)+\dfrac{iq\mu}{a(t)}\Delta_{\nu}^{T}(q,\mu,t)-4 K\dfrac{q}{a(t)}\dfrac{\partial}{\partial \mu}\Delta_{\nu}^{T}(q,\mu,t)=-2 \dot{D}_{q}(t)~~~~~~~~~~~~
\end{eqnarray}
 We can find a solution for the above equation with Green function method as:
\begin{equation}
\Delta_{\nu}^{T}(q,\mu,\tau)=\int d\tau d\mu^{\prime}G(\mu,\mu^{\prime},\tau,\tau^{\prime}) (-2\dot{D}_{q}(\tau^{\prime}))
\end{equation}
\section{Damping effect in open de-Sitter spacetime}
The Green function is obtained in open universe, $K=-1$, from Appendix (for more details, please see Appendix A).
\begin{eqnarray}
G(\mu,\mu^{\prime},\tau,\tau^{\prime})=\dfrac{i}{2 \pi }\Theta(\tau^{\prime}q-\tau q) e^{-i\tau q(\mu+2 \tau q)}e^{i\tau^{\prime} q(\mu^{\prime}+ 2\tau^{\prime}q)}
\end{eqnarray}
In this case, $\Delta_{\nu}(q,\mu,\tau)  $ is
\begin{eqnarray}
\Delta_{\nu}(q,\mu,\tau)= \dfrac{i}{2 \pi }e^{-i\tau q(\mu- 2 \tau q)}\int _{-1}^{+1}d\mu^{\prime} \int_{0}^{\pi} d\tau^{\prime}[-2\dot{D}_{q}(\tau^{\prime})] \Theta(\tau^{\prime}q-\tau q) e^{i\tau^{\prime} q(\mu^{\prime}- 2\tau^{\prime}q)}\nonumber\\
\end{eqnarray}
In the tensor mode, the only nonvanishing component is $ \delta T^{i}_{~\nu j} $:
\begin{eqnarray}
\delta T^{i}_{~\nu j}(\vec{x},t)&=&a^{-4}(t) \int d^{3}p ~\delta n_{\nu}(\vec{x},\vec{p},t)p \hat{p}_{i}\hat{p}_{j}\nonumber\\ &=&\bar{\rho}_{\nu}(t)\Sigma_{\lambda}  \int d^{3}q \beta (\vec{q},\lambda) e^{i\vec{q}.\vec{x}}e_{ij}(\hat{q},\lambda)\times \frac{1}{4}\int\frac{d^{2}\hat{p}}{4\pi}\Delta_{\nu}^{T}(q,\hat{p}.\hat{q},t)(1-(\hat{p}.\hat{q})^{2})^{2}
\end{eqnarray}
This is the neutrino contribution of the anisotropic inertia tensor $ \Pi^{T}_{ij} $:
\begin{eqnarray}
\Pi^{T}_{ij}=\frac{\bar{\rho}_{\nu}(t)}{4}\int\frac{d^{2}\hat{p}}{4\pi}\Delta_{\nu}^{T}(q,\hat{p}.\hat{q},t)(1-(\hat{p}.\hat{q})^{2})^{2}
\end{eqnarray}
With $ \hat{p}.\hat{q}=\mu$, the anisotropic inertia tensor $ \Pi^{T}_{ij} $ will be: 
\begin{eqnarray}
\Pi^{T}_{ij}&=&-\dfrac{\bar{\rho}_{\nu}(t)}{4}(\dfrac{1}{2\pi })e^{2i\tau^{2} q^{2}} \dfrac{1}{\tau^{5} q^{5}}[(-16\tau^{2} q^{2}+48)\sin(\tau q)-48\tau q \cos(\tau q)] \nonumber\\&&~~~~~~~~~~~~ \times \int _{-1}^{+1}d\mu^{\prime} \int_{0}^{\tau q} d\tau^{\prime}[-2\dot{D}_{q}(\tau^{\prime})] e^{-i\tau^{\prime} q(\mu^{\prime}+ 2\tau^{\prime}q)}
\end{eqnarray}
On the right hand side, the second and third terms are too smaller than the first term, so by ignoring them, the integrodifferential equation of gravitational waves in the presence of inertia tensor of neutrinos becomes:
\begin{eqnarray}\label{7}
\ddot{D}_{n}(t)+3\dfrac{\dot{a}}{a}D_{n}(t)+\dfrac{q^{2}}{a^{2}(t)}D_{n}(t)=-64\pi G\bar{\rho}(\tau)e^{2i\tau^{2} q^{2}}\dfrac{16}{\pi q} \dfrac{\sin \tau q}{\tau^{3} q^{3}} \int_{0}^{\tau q} d\tau^{\prime}[\dot{D}_{q}(\tau^{\prime})]\dfrac{\sin \tau^{'}q}{\tau^{'}q} e^{-2i\tau^{\prime 2} q^{2}}\nonumber\\
\end{eqnarray}
The right hand side of the equation (\ref{7}) is more complicated than the case of flat spacetime\cite{damping}. In the next section, we will calculate the decay of gravitational waves in the radiation dominated era.
\section{SHORT WAVELENGTHS at Radiation dominated era}
To investigate a short enough wavelength to have re-entered the horizon during radiation dominated era, we consider a new parameter as  $y=\dfrac{\bar{\rho}_{M}}{\bar{\rho}_{R}}=\dfrac{a}{a_{EQ}} $, where $a_{EQ}$ is the value of the Robertson-Walker scale factor at the matter radiation. In general, the fraction of the total energy density in neutrinos is:
\begin{eqnarray}
f_{\nu}(y)=\frac{\Omega_{\nu}(\frac{a_{0}}{a})^{4}}{\Omega_{M}(\frac{a_{0}}{a})^{3}+\Omega_{R}(\frac{a_{0}}{a})^{4}+\Omega_{\Lambda}}=\frac{f_{\nu}(0)}{1+y+(\frac{\Omega_{\Lambda}\Omega_{R}^{3}}{\Omega_{M}^{4}})y^{4}}
\end{eqnarray}
In which    $f_{\nu}(0)=\frac{\Omega_{\nu}}{\Omega_{\nu}+\Omega_{\gamma}}=0.40523 $. In the radiation dominated era, $y=\dfrac{\bar{\rho}_{M}}{\bar{\rho}_{R}}\ll 1$. So, the second and third terms in the denominator are too small and can be ignored. Also, in the presence of the anisotropic inertia tensor with change $ u=q \tau=q \int_{0}^{t} \dfrac{dt}{a(t)}=\dfrac{2qt}{a(t)}$, instead of $ t $, and using the Friedmann equation $ \dfrac{8\pi G \bar{\rho}}{3}=H^{2}=\dfrac{1}{4 t^{2}} $, the gravitational wave equation(\ref{7}) in radiation dominated era becomes:
\begin{eqnarray}\label{1}
\dfrac{d^{2}}{d u^{2}}D_{n}(u)+(\dfrac{2}{u})\dfrac{d}{du}D_{n}(u)+D_{n}(u)=\dfrac{-384 f_{v}}{\pi u^{2}} e^{2iu^{2}}\dfrac{\sin u}{u^{3}}\int_{0}^{u}e^{-2iu^{'2}}\dfrac{\sin u^{'}}{u^{'}}\dfrac{d D_{n}(u^{'})}{d u^{'}}du^{'}
\end{eqnarray}
when $ u\ll 1 $, the homogeneous solution is the constant value of $ D_{n}^{0} $; then, $ \dfrac{d D_{n}(u^{'})}{d u^{'}}=0$. So, the gravitational waves equation becomes:
\begin{eqnarray}
\dfrac{d^{2}}{d u^{2}}D_{n}(u)+(\dfrac{2}{u})\dfrac{d}{du}D_{n}(u)+D_{n}(u)=0
\end{eqnarray}
But, for $ u\gg 1 $, the homogeneous solution is  $ D_{n} = \dfrac{\sin u}{u} $; therefore, instead of 
$ \dfrac{d D_{n}(u^{'})}{d u^{'}}$ to the right hand side of the equation, we used: $ \dfrac{u^{'} \cos u^{'}-\sin u^{'}}{u^{'^{2}}}$. 
Therefore, the integro-differential equation (\ref{1}) becomes:
\begin{eqnarray}\label{8}
\dfrac{d^{2}}{d u^{2}}D_{n}(u)+(\dfrac{2}{u})\dfrac{d}{du}D_{n}(u)+D_{n}(u)=\dfrac{-384 f_{v}}{\pi u^{2}} e^{2iu^{2}}\dfrac{\sin u}{u^{3}}\int_{0}^{u} du^{'} e^{-2iu^{'2}} \dfrac{\sin u^{'}}{u^{'}}\dfrac{u^{'} \cos u^{'}-\sin u^{'}}{u^{'^{2}}}
\end{eqnarray}
For the integral, we used part-by-part method:  
\begin{eqnarray}
\int_{0}^{u}e^{-2iu^{'2}}\dfrac{\sin u^{'}}{u^{'}}\dfrac{u^{'} \cos u^{'}-\sin u^{'}}{u^{'^{2}}}=\dfrac{\sin^{2}u}{2 u^{2}}e^{-2iu^{2}}+2 i\int_{0}^{u}du^{'}\dfrac{\sin^{2}u^{'}}{2 u^{'}} e^{-2iu^{'2}}
\end{eqnarray}
in $ u\gg 1 $, $ \int_{0}^{u}du^{'}\dfrac{\sin^{2}u^{'}}{2 u^{'}} e^{-2iu^{'2}}=0 $; thus,
\begin{eqnarray}\label{101}
\dfrac{d^{2}}{d u^{2}}D_{n}(u)+(\dfrac{2}{u})\dfrac{d}{du}D_{n}(u)+D_{n}(u)=-\dfrac{35.75 f_{v}(0)}{\pi}\dfrac{\sin^{3} u}{u^{7}}
\end{eqnarray}
and its general solution is:
\begin{eqnarray}
D_{n}(u)=[D_{n}^{0} +\dfrac{32}{240}\dfrac{35.75 f_{v}(0)}{\pi}Ci(2u)+\dfrac{256}{240}\dfrac{35.75 f_{v}(0)}{\pi}Ci(4u)]\dfrac{\sin u}{u}+ D_{n}^{1}~\dfrac{35.75 f_{v}(0)}{24\pi}\dfrac{\sin 2u}{u^{2}}\nonumber\\
\end{eqnarray}
where $ D_{n}^{0} $ and $ D_{n}^{1} $ are constant and $ Ci(x) $ is the Cosine integral as $ Ci(x) =\gamma +\ln(x) +\int_{0}^{x}\dfrac{\cos t -1}{t} dt $. Deep inside of horizon when $ u\gg 1 $, the right hand side of Eq. (\ref{101}) becomes negligible and the solution approaches a homogeneous solution as:
\begin{eqnarray}
D_{n}(u)\longrightarrow \frac{\sin u}{ u}
\end{eqnarray}
For large $ u ~(u\gg 1)$,  $ Ci(2u) $ and $ Ci(4u) $ tend to zero. So, $ D_{n}^{0}=1 $ and $ D_{n}^{1}=0 $. Also, a numerical solution of Eq.(\ref{101}) shows that $ D_{n}(u) $ follows the $ f_{v}(0) $ solution pretty accurately until $ u \approx 1 $ when the perturbation enters the horizon [as compared with the solution $ \dfrac{\sin u}{u} $ for $ f_{v}(0) $]. Thereafter, the solution rapidly approaches the $ 0.5664 \dfrac{\sin u}{u} $, because $Ci(2)= 0.422981$ and  $Ci(4)= -0.140982$. \\Therefore, the effect of neutrino damping reduces the tensor amplitude by the factor of $0.5664 $ in the open cosmology, while in the flat and closed cases, the factor is $ 0.8026 $ and $0.4910$, respectively\cite{damping, Khodagholizadeh:2014ixa}. Hence, the tensor contribution to the temperature multipole coefficient $ C_{l} $ and the whole $ ^{''}B-B^{''} $  polarization multipole coefficient  $ C_{lB}$ will be $ 67.9\% $ less than they would be without damping due to free-streaming neutrinos.\\ Therefore, in open cosmology, the amplitude of the gravitational waves in the radiation dominated era and in the presence of neutrinos will be less than the flat case and can be more than closed cosmology, as shown in Fig.\cite{1}. But analytically, it can be shown as a ratio of two amplitude of gravitational waves as:
\begin{equation}
D_{n}^{open}(u)=1.15D_{n}^{closed}(u)
\end{equation}
In other words, the effect of neutrinos on the gravitational waves damping in open spacetime is $15\%$ more than in the closed case; in the absence of neutrinos, they are equal.      
\begin{figure}
	\includegraphics[scale=0.7]{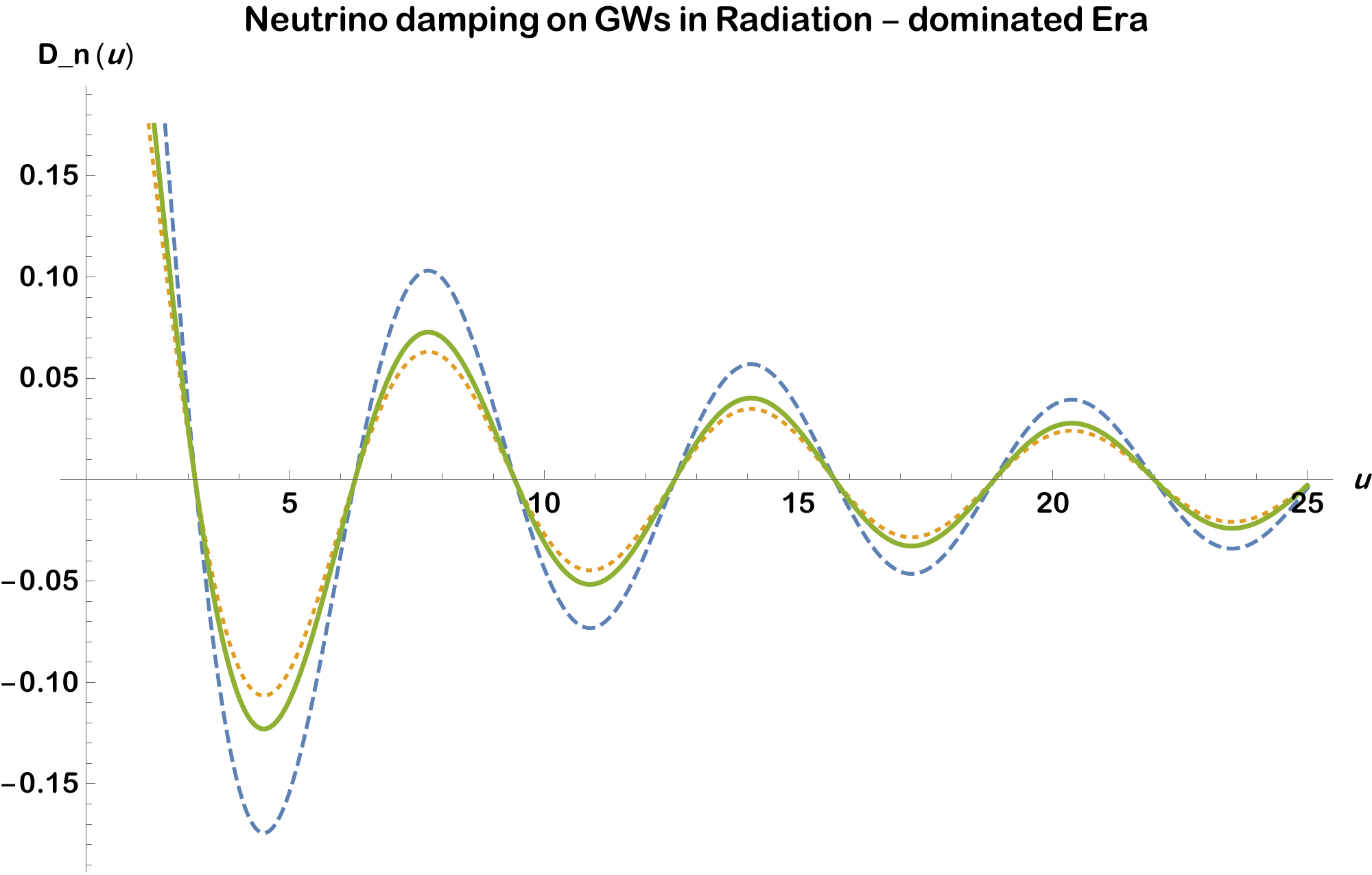}
	\caption{In radiation dominated era when the perturbation enters the horizon,  the free-streaming neutrinos in closed cosmology (dotted line-red) have a greater effect than the open case (solid line-green) on the damping of gravitational waves. Neutrinos damping in curved de Sitter spacetime is greater than the flat case.}
	\label{fig:1}
\end{figure}
\section{Damping in the matter dominated era}
Similar to the previous section, we will apply this method in matter and $ \Lambda$ dominated eras. The homogeneous solution of tensor mode in matter dominated is $j_{1}(u)$  or $\dfrac{\sin u}{u^{2}}$ which for nonzero  $f_{v}(0)$, the solutions approach  $0.9861\dfrac{\sin u}{u^{2}}$ and  $ 0.626\dfrac{\sin u}{u^{2}}$ with high accuracy  for flat and closed spacetime, respectively \cite{Khodagholizadeh:2017ttk}. The point is GWs decays as $\dfrac{1}{u}$ from radiation to matter era, where $u\approx \dfrac{qt}{a(t)}$. The integral-differential Eq.(\ref{1}) in matter dominated spacetime becomes:
\begin{eqnarray}
\dfrac{d^{2}}{du^{2}}D_{n}(u)+\dfrac{4}{u}\dfrac{d}{du}D_{n}(u) +D_{n}(u)=-\dfrac{35.75 f_{\nu}(0)}{\pi} \dfrac{\pi}{6} \dfrac{\sin u}{u^{5}}
\end{eqnarray}
The general solution is: 
\begin{eqnarray}
D_{n}(u)= (D_{n}^{0}-\dfrac{35.75 f_{\nu}(0)}{36}\dfrac{\sin^{2}u}{u})\dfrac{\sin u}{u^{2}}
\end{eqnarray}
When $u\gg 1$, $ D_{n}(u)$ tends to be zero, so the constant coefficient will be $ D_{n}^{0}=1 $. Also, in condition $ u \approx 1 $ when the perturbation enters the horizon (as compared with the answer $ \dfrac{\sin u}{u^{2}} $ for $ f_{v}(0)=0 $), the solution rapidly approaches $ 0.8963 \dfrac{\sin u}{u^{2}} $. Therefore, in the matter-dominated era, the effect of neutrinos reduces the tensor amplitude by a factor of $ 0.8963 $ in opened cosmology, while the factors are $ 0.9860 $ and $0.6258$ in flat and close universe, respectively. The tensor contribution to the   $ ^{''}\verb"B-B"^{''} $ polarization multipole coefficient $ C_{lB}$ will be $ 19.6 \% $ less than what they would be without damping, as the result of free-streaming neutrinos in open spacetime.  Consequently, in open cosmology, neutrinos have a greater effect on gravitational waves damping compared to the flat case, but still less than closed cosmology, as shown in Fig.\ref{fig2}.
\section{ General wavelengths and  $\Lambda-$ dominated era}
To investigate the tensor perturbation entering horizon after the vacuum energy becomes important, it is convenient to change the independent variable  $~ t~ $  to  $ \chi \equiv \frac{\bar{\rho}_{_{\Lambda}}}{\bar{\rho}_{_{M}}}=\frac{\bar{\rho}_{_{\Lambda,EQ}}}{\bar{\rho}_{_{M,EQ}}}\dfrac{a^{3}}{a_{_{EQ}}^{3}} $, where $ a_{_{EQ}} $ ,  $\bar{\rho}_{_{\Lambda,EQ}} $ and $ \bar{\rho}_{_{M,EQ}} $ are the values of the Robertson-Walker scale factor, the energy densities of vacuum, and  matter at equality of matter-$ \Lambda $ era. From the Friedmann equation, with $\Omega_{K}=\dfrac{1}{a^{2} H^{2}}$, we have:
\begin{eqnarray}\label{10}
H_{EQ}\dfrac{dt}{\sqrt{2}}=\dfrac{d \chi}{3\sqrt{(1+\Omega_{K,EQ})(\chi+\chi^{2})+2\Omega_{K,EQ}\chi^{4/3}}}
\end{eqnarray}
where $ H_{EQ} $ and $ \Omega_{_{K,EQ}} $  are the expansion rate and the curvature density at matter-vacuum equality, respectively. In $ \Lambda $ dominated era, we take $ \chi\gg 1 $; so,
\begin{figure}
	\includegraphics[scale=0.7]{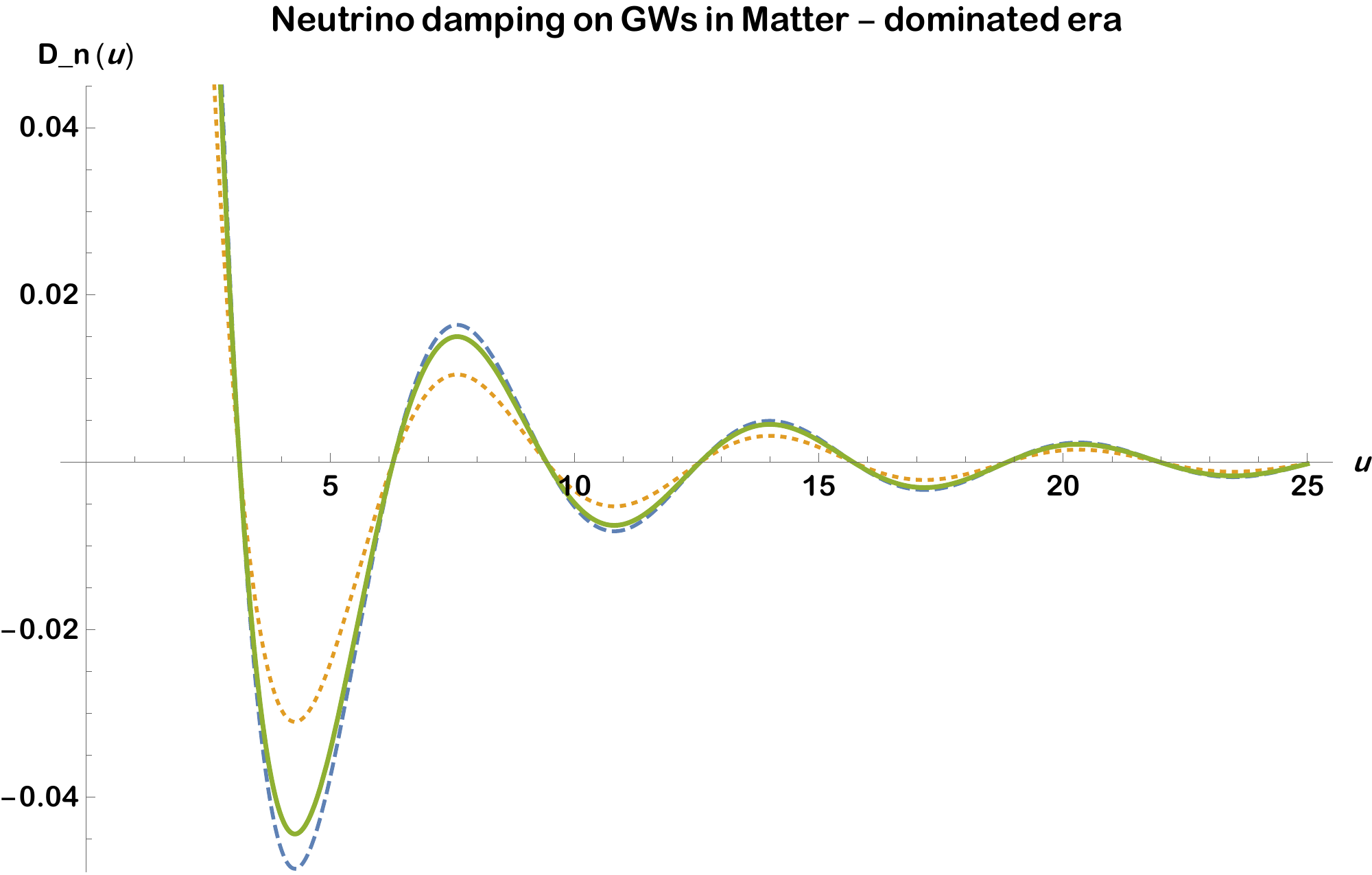}
	\caption{In a matter-dominated era, the damping of the free-streaming neutrinos on GWs in open cosmology (solid line-green) is slightly greater than the flat case(dashed line-blue) and less than the closed cosmology (dotted line-red). It seems that the ratio of neutrino effect in reducing the amplitude of GWs in the matter-dominant the same as the radiation dominant era.}
	\label{fig2}
\end{figure}
\begin{eqnarray}
H_{EQ}\dfrac{dt}{\sqrt{2}}=\dfrac{dX}{3\chi\sqrt{1+\Omega_{K,EQ}}}
\end{eqnarray}
As mentioned, $ \Omega_{K,EQ} $ is the curvature density when the matter and vacuum densities are equal. Therefore, the filed equation becomes:
\begin{eqnarray}\label{110}
\dfrac{d^{2}}{d\chi^{2}}D_{n}(\chi)+\dfrac{2}{\chi}\dfrac{d}{d\chi}D_{n}(\chi)+\dfrac{\xi^{2}}{\chi^{8/3}}D_{n}(\chi)=0
\end{eqnarray}
where $ \xi=\dfrac{\sqrt{2}~n}{3 H_{EQ}~a_{EQ}(1+\Omega_{K,EQ})^{1/2}} $  is the dimensionless rescaled wave number, while $ n $ is discrete number for curved spacetime. Whatever the value of $ \xi $ , the general solution is:
\begin{eqnarray}
D_{n}(\chi)=D_{n}^{0} [(\dfrac{3\xi}{\chi^{1/3}})\sin(\dfrac{3\xi}{X^{1/3}})+\cos(\dfrac{3\xi}{\chi^{1/3}})]
\end{eqnarray}
where $D_{n}^{0} $ is constant. When $ \dfrac{n}{a}= \dfrac{\xi}{\chi^{1/3}}  $  is very large relative to the Hubble parameter, the perturbations are deep inside the horizon and the solution will be:
\begin{eqnarray}
D_{n}(\chi)=D_{n}^{0} (\dfrac{3\chi}{\chi^{1/3}})\sin(\dfrac{3\chi}{\chi^{1/3}}).
\end{eqnarray}   
All wavelengths are multiplied by the effect of neutrinos  in the $\Lambda-$ dominated era as: 
\begin{eqnarray}\label{42}
D_{n}(\chi)=\alpha (\xi) D_{n}^{0} (\dfrac{3\xi}{\chi^{1/3}})\sin(\dfrac{3\xi}{\chi^{1/3}})
\end{eqnarray}
where $\alpha (\xi)$ is $ 0.8963$ for $ \xi\gg 1$ and $\alpha (\xi)=1$ for $\xi\ll 1$, because the damping effect is small anyway for $\xi\ll 1$. Therefore, it will be an adequate approximation for all the wavelengths to take the gravitational wave amplitude in  $\Lambda$ dominated era to be given by multiplying the result by a factor  $\alpha (\xi)\cong \dfrac{1+ 0.8963 \xi}{1+\xi }$ and all the observable effects of cosmological gravitational waves will be reduced slightly by this factor. 
It is noticed that we ignore a shift in phase of the oscillation for longer wavelength, because the amplitude of gravitational waves at the time of decoupling is a sensitive function of wavenumber, but the maximum effect is roughly $5\%$ in $\xi=O(1)$.\\ For all the wavelengths, we have:
\begin{eqnarray}\label{33}
\dfrac{D_{n}^{open}(\chi)}{D_{n}^{close}(\chi)}&=& \dfrac{\dfrac{1+ 0.8963 \xi}{1+ \xi}}{\dfrac{1+0.6258\kappa}{1+\kappa} } \dfrac{ \xi\sin(\dfrac{3\xi}{\chi^{1/3}})}{\kappa\sin(\dfrac{3\kappa}{\chi^{1/3}})} \nonumber\\
&\cong&\frac{7.1908 H_{EQ}^2a_{EQ}^2 \sqrt{1-\Omega_{K,EQ}^2}+6.42803 H_{EQ}~a_{EQ} n \sqrt{1+\Omega_{K,EQ}}+1.43225 n^2}{7.1908 H_{EQ}^2 a_{EQ}^2 \sqrt{1-\Omega_{K,EQ}^2}+5.51042 H_{EQ}~a_{EQ} n \sqrt{1-\Omega_{K,EQ}}+ n^2} \sqrt{\dfrac{1-\Omega_{K,EQ}}{1+\Omega_{K,EQ}}}\nonumber\\
\end{eqnarray}
where $ \kappa=\dfrac{\sqrt{2}~n}{3 H_{EQ}~a_{EQ}(1-\Omega_{K,EQ})^{1/2}} $ \cite{Khodagholizadeh:2014ixa}. At first, we consider the case of perturbation inside the horizon, i.e., for which $n/a\gg H_{EQ}$, we have:
\begin{eqnarray}\label{deep}
\dfrac{D_{n}^{open}(\chi)}{D_{n}^{close}(\chi)}=1.43  \sqrt{\dfrac{1-\Omega_{K,EQ}}{1+\Omega_{K,EQ}}}
\end{eqnarray} 
By considering the inter media range, where  $n/a\gg H_{EQ}$ and $ \chi \ll 1$, this ratio will be:
\begin{eqnarray}
\dfrac{D_{n}^{open}(\chi)}{D_{n}^{close}(\chi)}\cong 1.16 
\end{eqnarray}  
In the expanding universe, $ H_{EQ} $ is important and the universe enters the $\Lambda-$ dominated era; there is no difference between closed and open de sitter spacetime. However, the amplitude of gravitational waves in open de Sitter spacetime is greater than the closed case, while it is in contrast to the neutrino effect on their damping but anyway it could be stable from the outside horizon to the deep inside. After all, it is relative to the value of curvature density when matter and vacuum densities are equal, although it is very small relative to the other densities.\\
 But the highest neutrinos effect on GWs is seen in $\xi\gg 1$, which increases this difference by up to $43\%$; however, in later times, it is up to $16\%$ and, in the present era, there will be no difference.  The existing ground-based operators, such as LIGO and VIRGO at about 100 Hz,  reported GWs that are coming from black-hole mergers \cite{GWs}. These detectors do not have the sensitivity to detect cosmological GWs; thus, they might be detected by space-borne laser interferometers, which operate at frequencies around $0.01$ to $0.1$ Hz. Therefore, a GW with a frequency of $qc/2\pi a_{0}=10^{-2}  $ Hz  \cite{Komatso} would have $\xi\cong 1.3 \times 10^{15}/\Omega_{M}h^{2} \gg 1$ at the present epoch which it is compatible with the condition of maximum neutrino effect on GWs.\\
\section{CONCLUSION}
We derived the integrodifferential equation of gravitational waves with free-streaming neutrinos as a part of the anisotropic stress tensor in spatially open de Sitter spacetime. The solution showed that the presence of neutrinos reduced the amplitude by a factor $0.5664$ and $0.8963$ in radiation and matter-dominated eras, respectively and in the next era, the amplitude would be less than before. Also, the $"B-B"$ polarization multipoles coefficients $C_{lB}$, are $ 67.9\%$ and  $ 19.6\%$ were less than those without the damping due to neutrinos and anti-neutrinos.
The other point was the result which depended on the $ \Omega_{k}$. When $\Omega_{k}\longrightarrow 0$, the solution did not approach the flat case thus there was a topological effect coming from a general topology.\\
Although in the radiation dominated era the density of neutrinos has a greater effect on gravitational wave damping, this effective difference is less than the next era. Also this effect, regardless of the spacetime curvature, had a decreasing trend from after inflation until now.  Therefore, we can introduce neutrinos  as a new tool to measure the spatial curvature of the universe by using future space-borne laser interferometers such as the Big Bang Observer and the Decihertz Laser Interferometer Gravitational Waves Observatory.
\section*{Appendix A: The Green function}
At first we rewrite the Eq.(\ref{12}) as
\begin{eqnarray}
\dfrac{\partial}{\partial t}\Delta_{\nu}^{T}(q,\mu,t)+\dfrac{iq\mu}{a(t)}\Delta_{\nu}^{T}(q,\mu,t)-4 K\dfrac{q}{a(t)}\dfrac{\partial}{\partial \mu}\Delta_{\nu}^{T}(q,\mu,t)=-2 \dot{D}_{q}(t)~~~~~~~~~~~~
\end{eqnarray}
A general solution of above equation with Green function method can  be written as 
\begin{equation}
\Delta_{\nu}^{T}(q,\mu,\tau)=\int d\tau d\mu^{\prime}G(\mu,\mu^{\prime},\tau,\tau^{\prime}) (-2\dot{D}_{q}(\tau^{\prime}))
\end{equation}
With $K=-1$ and  by defining $q^{\prime}=\dfrac{q}{a(t)}$  as a ratio of  wave number to scale factor we have
\begin{eqnarray}
\dfrac{\partial}{\partial t}\Delta_{\nu}^{T}(q,\mu,t)+i q^{\prime} \mu\Delta_{\nu}^{T}(q,\mu,t)+ 4 Kq^{\prime}\dfrac{\partial}{\partial \mu}\Delta_{\nu}^{T}(q,\mu,t)=\lambda\Delta_{\nu}^{T}(q,\mu,t)
\end{eqnarray} 
or
\begin{eqnarray} \label{42}
\dfrac{\partial}{\partial t}\Delta_{\nu}^{T}(q,\mu,t)+(i q^{\prime} \mu-\lambda)\Delta_{\nu}^{T}(q,\mu,t)+ 4 Kq^{\prime}\dfrac{\partial}{\partial \mu}\Delta_{\nu}^{T}(q,\mu,t)=0
\end{eqnarray} 
we can express the solution as 
\begin{eqnarray}\label{43}
\Delta_{\nu}^{T}(q,\mu,t)=\int d\omega \int d\Omega ~e^{i\omega T}~e^{i\mu \Omega}g(\omega, \Omega)
\end{eqnarray}
by substitute (\ref{43}) in (\ref{42}) we have
\begin{eqnarray}
\int d\omega \int d\Omega   \{ (i\omega-\lambda) g(\omega, \Omega) -q^{\prime} \dfrac{\partial}{\partial \Omega}g(\omega, \Omega) + 4i q^{\prime} \Omega  g(\omega, \Omega) \} ~e^{i\omega T}~e^{i\mu \Omega}=0
\end{eqnarray}
or
\begin{eqnarray}
(i\omega-\lambda + 4i q^{\prime} \Omega ) g(\omega, \Omega) -q^{\prime} \dfrac{\partial}{\partial \Omega}g(\omega, \Omega)=0 
\end{eqnarray} 
above equation has a simple solution as 
\begin{eqnarray}
g(\omega, \Omega)= C_{1} e^{[\dfrac{i\omega-\lambda}{q^{\prime}}\Omega+2 i \Omega^{2}]}
\end{eqnarray}
Therefore (\ref{43}) will be 
\begin{eqnarray}
\Delta_{\nu}^{T}(q,\mu,t)&=&\int d\omega \int d\Omega ~e^{i\omega T}~e^{i\mu \Omega}e^{[\frac{i\omega-\lambda}{q^{\prime}}\Omega+2 i \Omega^{2}]}\nonumber\\ &=&\int d\omega \int d\Omega e^{i\omega (T+\frac{\Omega}{q^{\prime}})} e^{i\Omega(\mu+2\Omega)} e^{-\frac{\lambda}{q^{\prime}}\Omega}
\end{eqnarray}
As a generally the Green function will be 
\begin{eqnarray}
G(\mu,\mu^{\prime},\tau,\tau^{\prime})=\int  \dfrac{ d\lambda}{\lambda}G(\mu,\tau,\lambda) G^{\ast}(\mu^{\prime},\tau^{\prime},\lambda)
\end{eqnarray}
Thus
\begin{eqnarray}
G(\mu,\mu^{\prime},\tau,\tau^{\prime})&=&\int d\omega~ d\Omega\int d\omega^{\prime}~d\Omega^{\prime}\int \dfrac{d\lambda}{\lambda}  e^{i\omega (\tau+\frac{\Omega}{q^{\prime}})} e^{i\Omega(\mu+2\Omega)} e^{-\frac{\lambda}{q^{\prime}}\Omega}e^{-i\omega^{\prime} (\tau^{\prime}+\frac{\Omega^{\prime}}{q^{\prime}})} e^{-i\Omega^{\prime}(\mu^{\prime}+2\Omega^{\prime})} e^{-\frac{\lambda}{q^{\prime}}\Omega^{\prime}}\nonumber\\
\end{eqnarray}
By using the define the $ \Theta(\Omega+\Omega^{\prime})=\int \frac{d\lambda}{\lambda} e^{-\frac{\lambda}{q^{\prime}}(\Omega+\Omega^{\prime})} $, the Green function will be 
\begin{eqnarray}
G(\mu,\mu^{\prime},\tau,\tau^{\prime})&=&\int d\omega~ d\Omega\int d\omega^{\prime}~d\Omega^{\prime} \Theta(\Omega+\Omega^{\prime}) e^{i\omega (\tau+\frac{\Omega}{q^{\prime}})} e^{i\Omega(\mu+2\Omega)} e^{-i\omega^{\prime} (\tau^{\prime}+\frac{\Omega^{\prime}}{q^{\prime}})} e^{-i\Omega^{\prime}(\mu^{\prime}+2\Omega^{\prime})} \nonumber\\
\end{eqnarray}
Now by using the  below $\delta-$functions
\begin{eqnarray}
\int d\omega e^{i\omega (\tau+\frac{\Omega}{q^{\prime}})}&=&(2\pi)^{3}\delta (\tau+\frac{\Omega}{q^{\prime}})\nonumber\\
\int d\omega^{\prime} e^{i\omega^{\prime} (\tau^{\prime}+\frac{\Omega^{\prime}}{q^{\prime}})}&=&(2\pi)^{3}\delta (\tau^{\prime}+\frac{\Omega^{\prime}}{q^{\prime}})
\end{eqnarray}
In the last: 
\begin{eqnarray}
G(\mu,\mu^{\prime},\tau,\tau^{\prime})=\dfrac{1}{2\pi}\Theta(\tau^{\prime}q-\tau q) e^{-i\tau q(\mu-2\tau q)}~e^{i\tau^{\prime} q(\mu^{\prime}-2\tau^{\prime} q)}
\end{eqnarray}
Which is the Green function in open de Sitter spacetime.

\end{document}